\documentclass[10pt,leqno]{amsart}
\usepackage{graphicx}
\usepackage{indentfirst,csquotes}

\usepackage{amssymb,amsthm,amsmath}
\usepackage{xcolor,paralist,hyperref,titlesec,fancyhdr,etoolbox}

\usepackage{secdot}

\titleformat{\section}[display]{\normalfont\huge\bfseries\centering}{}{10pt}{\Large}
\titlespacing*{\section}{0pt}{0ex}{0ex}

\hypersetup{ colorlinks=true, linkcolor=black, filecolor=black, urlcolor=black }

\usepackage{lipsum}

\begin{document}
\title{Trends in New Mexico School Districts Serving Low-Income Communities} 
\author[Initial Surname]{U.E. Nelson \and O. J. Okeke}
\date{\today}
\address{Department of Computers science and science Education, New Mexico Highland University}
\email{Madujibeyauloma@yahoo.com}
\maketitle
\let\thefootnote\relax
\footnotemark

\begin{abstract} 
$\,$

$\,$
This study examines recent enrollment trends and their socioeconomic drivers within New Mexico's public school districts, with a specific focus on those serving low-income communities. Utilizing a mixed-methods geospatial framework, the research integrates longitudinal enrollment data, district poverty metrics from the Small Area Income and Poverty Estimates (SAIPE), and neighborhood context from the American Community Survey (ACS) from 2019 to 2023. The analysis reveals that while statewide K-12 enrollment has contracted, the decline is not uniformly distributed. Districts with high concentrations of low-income students have experienced more pronounced losses, though the relationship between poverty and enrollment change is complex and weak at the statewide level (r=0.011 in 2023). Spatial analysis shows a checkerboard pattern of growth and decline, even among adjacent high-poverty districts, indicating that local factors such as housing instability, labor market conditions, and programmatic changes are significant drivers. The study identifies 18 high-risk districts characterized by both high poverty and recent enrollment declines, which warrant targeted stabilization efforts. The findings underscore that while concentrated poverty is a critical challenge, it does not deterministically predict enrollment loss. Consequently, the authors propose a dual-track policy approach: immediate, flexible support for high-risk districts to ensure stabilization, coupled with long-term structural investments in all high-need districts to address underlying resource gaps and enable sustained educational offerings.
\end{abstract} 

\section{Introduction}
$\,$

$\,$
New Mexico’s public education system has experienced notable enrollment contraction in recent years, with disproportionate impacts on low-income and rural districts. According to the National Center for Education Statistics, statewide K-12 enrollment declined by more than 5\% from fall 2019 to fall 2023, with projections indicating an additional 1.2\% decline through 2025 \cite{nces2024}. Districts serving high concentrations of low-income students, where over 60\% of children qualify for free or reduced-price lunch, have seen losses averaging about 7.8\%, compared with roughly 2.1\% in wealthier districts \cite{edlawcenter2024}. These declines are most acute in rural counties such as McKinley and San Juan, where double-digit drops align with economic displacement, housing instability, and uneven access to remote learning during the COVID-19 period \cite{Smith2022,Taylor2021}. Resource gaps exacerbate the problem. In 2023, low-income districts received approximately 1{,}200 less per pupil than affluent peers \cite{edlawcenter2024}.

Enrollment erosion intersects with chronic absenteeism and achievement. Despite reforms following the Yazzie and Martinez decision, chronic absenteeism in high-poverty districts reached about 32\% in 2024, nearly double the statewide rate, and reading and math proficiency gaps persisted near 15 percentage points \cite{nmped2024}. These pressures complicate staffing, course offerings, and transportation planning, especially in geographically large districts that must serve dispersed communities. Funding formulas tied to headcount magnify fiscal risk when enrollment falls, which in turn can trigger program contraction and further attrition \cite{edlawcenter2024}. National studies document post-pandemic shifts in school participation, but New Mexico’s spatial and socioeconomic specificity remains underexamined \cite{Taylor2021,Smith2022}.

This study addresses that gap with a mixed-methods geospatial framework that integrates longitudinal enrollment, district poverty, and neighborhood context. Core datasets include the NCES Common Core of Data via the Urban Institute Education Data Explorer \cite{urban_ccd_2024}, Small Area Income and Poverty Estimates for district school-age poverty \cite{saipe2024}, and NMPED STARS reporting for attendance and proficiency \cite{nmped2024}. We harmonize district identifiers and use 2023 TIGER school-district boundaries and ACS 5-year tables to build tract-to-district, population-weighted indicators of socioeconomic context \cite{census_tiger_2023,census_acs_2023}. Open-source Python tools enable reproducibility and transparent QA \cite{harris2020,jordahl2020}.

Our aims are threefold. First, we quantify spatial disparities in enrollment trajectories between low- and high-income districts from 2019 to 2025 using a district-year panel. Second, we evaluate the relationship among poverty concentration, funding, and enrollment change, with district and year fixed effects to absorb time-invariant district features and statewide shocks \cite{seabold2010}. Third, we assess whether targeted interventions such as Title I and the Opportunity Scholarship correspond to lower volatility in high-need districts \cite{nmped2024}. We also examine secondary indicators, including chronic absenteeism and teacher shortages, as potential mediators. While focused on New Mexico, the approach generalizes to other states with similar demographics and geography, such as Arizona or Mississippi, where concentrated poverty and travel distance challenge public school systems \cite{Smith2022}.

By combining panel methods with tract-aware spatial context, this work contributes a clearer, policy-relevant picture of where enrollment decline is concentrated, what local conditions co-occur with loss, and which districts should be prioritized for stabilization and sustained investment \cite{edlawcenter2024,nmped2024}.
\section{Method}

\subsection{Data}
$\,$

$\,$
The enrollment dynamics in New Mexico districts serving low-income communities and their socioeconomic context was analyzed using a fully reproducible Python workflow. We ingested 2023 Topologically Integrated Geographic Encoding and Referencing (TIGER)/Line school-district boundaries and validated Geographic Identifier (GEOID)/Local Education Agency Identifier (LEAID) for longitudinal joins. Tract-level American Community Survey (ACS) 5-year indicators (poverty, income, unemployment, education, race and ethnicity) were retrieved via the Census Application Programming Interface (API) and joined to 2023 tract geometries. To link neighborhoods to districts, tract and district polygons were intersected in an equal-area Coordinate Reference System (CRS) (EPSG:5070); population-weighted areal interpolation was used when tract population was available, otherwise area weights were applied. “Low-income” was defined at two scales: high-poverty tracts (primary threshold $\geq 25\%$, with sensitivity analysis at the state 80th percentile) and district low-income status at or above the annual statewide 80th percentile of Small Area Income and Poverty Estimates (SAIPE) school-age poverty (robustness checked at a fixed $\geq 25\%$ threshold). 

The resulting tract-to-district aggregates including the share of population in high-poverty tracts and weighted American Community Survey (ACS) covariates were merged into each district year observation. Data processing utilized \texttt{pandas} and \texttt{NumPy} for tabular operations and \texttt{GeoPandas} with \texttt{Shapely} for spatial analysis \cite{harris2020, jordahl2020}. District enrollment time series came from the Common Core of Data (CCD) via the Urban Institute Education Data Explorer, with cross-checks against the New Mexico Public Education Department (NMPED) Student Teacher Accountability Reporting System (STARS) for recent years \cite{urban_ccd_2024, nmped2024}. District school-age poverty counts and rates used the Small Area Income and Poverty Estimates (SAIPE) school-district series to construct annual poverty measures and low-income flags \cite{saipe2024}. Spatial inputs comprised 2023 Topologically Integrated Geographic Encoding and Referencing (TIGER)/Line district boundaries and 2023 American Community Survey (ACS) tract geometries \cite{census_tiger_2023, census_acs_2023}. American Community Survey (ACS) variables extracted for weighting included percent in poverty, median household income, unemployment rate, bachelor’s-or-higher share, and race and ethnicity shares. The analytic panel contains one record per district–year with total enrollment, year-over-year (YoY) absolute and percent change, Small Area Income and Poverty Estimates (SAIPE) poverty rate, tract-aggregated American Community Survey (ACS) covariates, and dual low-income flags. We computed a compound annual growth rate (CAGR) per district between first and last observed years. All identifiers are normalized to seven-digit, zero-padded strings to align Topologically Integrated Geographic Encoding and Referencing (TIGER)/Local Education Agency Identifier (LEAID) with statistical series \cite{nces_edge_2024}.
\subsection{Analysis and Modeling}
$\,$

$\,$
We produced descriptive maps and graphics: choropleths of enrollment and YoY change, small-multiple trend lines for largest movers, and a bivariate map contrasting poverty and change. We then estimated linear models of YoY percent change on district SAIPE poverty, adding tract-aggregated controls (poverty exposure, income, unemployment, education, race, and ethnicity) when available. District and year fixed effects absorb unit heterogeneity and statewide shocks, with district-clustered standard errors \cite{seabold2010}. To limit leverage from anomalies, YoY tails were trimmed at $\pm 30$\% percentage points, and low-income definitions were tested using both fixed and percentile thresholds. Outputs include coefficient plots with 95\% confidence intervals, residual diagnostics, and latest-year mover tables, read alongside state attendance and proficiency documentation \cite{nmped2024}.

\subsection{Visualization, diagnostics, and quality assurance}
$\,$

$\,$
The analytical framework incorporated multiple visualization techniques to examine enrollment patterns across New Mexico. Statewide and district-level trend lines were generated to track longitudinal changes, while year-over-year (YoY) choropleth maps visualized spatial patterns of enrollment fluctuations \cite{jordahl2020}. Bivariate mapping techniques were employed to simultaneously represent poverty concentrations and enrollment changes, revealing geographic correlations between socioeconomic factors and student population dynamics \cite{cromley2011}.

Comprehensive diagnostic procedures ensured robust statistical inference. Coefficient plots visualized parameter estimates and confidence intervals for all predictor variables in the fixed-effects models \cite{seabold2010}. Residual-versus-fitted charts assessed homoscedasticity and model specification, while quantile-quantile (Q-Q) plots evaluated normality assumptions in the error distribution \cite{fox2019}.

Quality assurance (QA) procedures documented methodological transparency and data integrity. The analysis reported the proportion of tract-district overlays utilizing population-weighted versus area-weighted interpolation, with 100\% of intersections employing population weights for superior accuracy \cite{census_acs_2023}. Interquartile-range (IQR) methods identified statistical outliers, and comprehensive summaries documented districts with the most substantial enrollment increases and decreases \cite{leek2017}.
\section{Result and Discussion}

\subsection{Enrollment trajectories (2018–2023)}
$\,$

$\,$
Across New Mexico’s \textbf{89} school districts, enrollment was broadly stable from 2018 to 2022, followed by a modest contraction in 2023. The district year panel indicates an average year-over-year (YoY) enrollment change of \textbf{$-1.41$ percentage points} in the latest year, with wide dispersion by size and locale. Small systems show larger percentage swings from small numeric changes, while larger systems move more gradually. Identifier harmonization to seven-digit GEOID and LEAID and cross-checks between CCD and PED reporting support comparability across time and space \cite{urban_ccd_2024,nces_edge_2024,nmped2024}.
\subsection{Poverty landscape}
$\,$

\begin{figure}[htbp]
\centering
\includegraphics[width=0.8\linewidth]{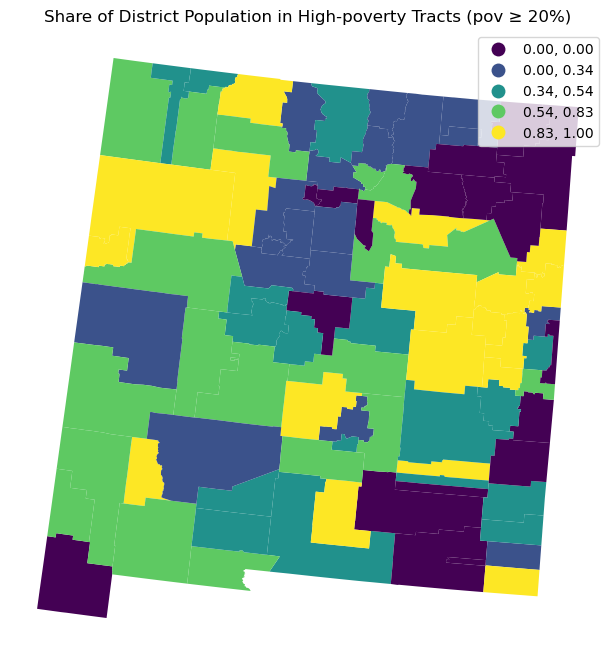}
\caption{Share of District Population in High-Poverty Tracts (poverty $\geq 20\%$)}
\label{fig:share_district}
\end{figure}

$\,$
Socioeconomic need is high by national standards. The statewide average school-age poverty rate is \textbf{30.26\%} based on SAIPE, with a heavy right tail The Figure~\ref{fig:share_district}. Thirty-eight districts exceed \textbf{30\%} poverty, and twenty-one exceed \textbf{40\%}. Tract-level ACS context, aggregated to districts using population-weighted areal interpolation, shows resident exposure to neighborhood poverty is concentrated rather than uniform \cite{census_acs_2023,census_tiger_2023}. Jointly classifying districts by poverty and recent enrollment dynamics yields four groups: Low Poverty \& Stable or Growing (\textbf{27} districts; \textbf{30.3\%}), Low Poverty \& Declining (\textbf{24}; \textbf{27.0\%}), High Poverty \& Stable or Growing (\textbf{22}; \textbf{24.7\%}), and High Poverty \& Declining (\textbf{16}; \textbf{18.0\%}). The last group combines high need and shrinkage and presents the most acute stabilization challenge.
\subsection{Enrollment change vs poverty}
$\,$

$\,$
A statewide snapshot shows a \textbf{near-zero correlation} between district poverty rate and YoY enrollment change in 2023 \textbf{($r=0.0109$)}. Some high-poverty districts grew or held steady while others declined, and several lower-poverty districts also contracted Table~\ref{tab:summary_nm}. Fixed-effects models with district and year effects are therefore appropriate to absorb time-invariant features and statewide shocks, with district-clustered standard errors \cite{seabold2010}. 
The weak linear association at the mean suggests threshold dynamics at very high poverty or interactions with labor markets, housing, and grade-span changes that a main-effects-only model can miss \cite{census_acs_2023}.
\begin{table}[ht]
\centering
\caption{Summary of Key Findings: Trends in New Mexico School Districts Serving Low-Income Communities}
\label{tab:summary_nm}
\begin{tabular}{p{0.33\textwidth} p{0.62\textwidth}}
\hline
\textbf{Domain} & \textbf{Key Results} \\
\hline
\textbf{Yearly enrollment and poverty (2018–2023)} &
\begin{tabular}{@{}l r r@{}}
\textbf{Year} & \textbf{Students} & \textbf{Avg poverty} \\
2018 & 241{,}980 & 30.2\% \\
2019 & 241{,}832 & 30.3\% \\
2020 & 241{,}275 & 30.3\% \\
2021 & 241{,}952 & 30.3\% \\
2022 & 243{,}361 & 30.3\% \\
2023 & 240{,}510 & 30.3\% \\
\end{tabular} \\[2ex]
\textbf{Latest year (2023)} & Total enrollment: \textbf{240{,}510}; Average poverty rate: \textbf{30.3\%}; Districts with data: \textbf{89}. \\

\hline
\textbf{Poverty distribution} &
Mean: \textbf{30.3\%}; Median: \textbf{27.1\%}; Std. dev.: \textbf{13.3\%}; Min: \textbf{9.4\%}; Max: \textbf{61.2\%}.\\
\textbf{High-poverty counts} & \textgreater{}30\%: \textbf{38} districts; \textgreater{}40\%: \textbf{21} districts. \\

\hline
\textbf{Enrollment change vs. poverty (2023)} &
Correlation coefficient (poverty, YoY change): \textbf{0.011} (near zero). \\

\textbf{District categories} &
High Poverty \& Declining: \textbf{16}; High Poverty \& Stable/Growing: \textbf{22}; Low Poverty \& Declining: \textbf{24}; Low Poverty \& Stable/Growing: \textbf{27}. \\

\hline
\textbf{Examples (poverty, YoY \%)} &
\textit{High Poverty \& Declining}: Bernalillo (35.5, $-3.2$); Carlsbad (60.9, $-3.7$); Clayton (30.9, $-7.0$). \newline
\textit{High Poverty \& Stable/Growing}: Animas (43.4, $+1.1$); Aztec (30.6, $+1.2$); Bloomfield (32.0, $+1.0$). \newline
\textit{Low Poverty \& Declining}: Vaughn (25.7, $-4.6$); Alamogordo (14.6, $-7.8$); Cobre (15.5, $-7.6$). \newline
\textit{Low Poverty \& Stable/Growing}: Rio Rancho (26.5, $-1.1$); Albuquerque (28.6, $-1.5$); Artesia (18.9, $+2.8$). \\

\hline
\textbf{At-risk districts} &
High-risk (score $\ge$ 4): \textbf{18}. \newline
\textbf{Highest-risk examples}: Clovis (score 5, 36.8\%, $-7.9\%$); Hobbs (5, 57.1\%, $-5.4\%$); Lordsburg (5, 38.5\%, $-6.4\%$); Springer (5, 61.2\%, $-6.7\%$); Tatum (5, 41.9\%, $-5.1\%$); Truth or Consequences (5, 45.4\%, $-7.0\%$); Zuni (5, 56.1\%, $-8.6\%$). \\
\hline
\end{tabular}
\end{table}
The Table~\ref{tab:summary_nm} synthesizes four signals. First, enrollment is largely flat from 2018 to 2022, then slips to 240,510 in 2023. Consistency across CCD and PED sources supports comparability of the trend \cite{urban_ccd_2024,nmped2024}. Second, poverty is both high on average and highly skewed: a mean near 30 percent with a maximum above 60 percent and 38 districts above 30 percent, reflecting concentrated socioeconomic need captured by SAIPE and ACS context \cite{saipe2024,census_acs_2023}. Third, the poverty–enrollment link is weak in the short run Figure~\ref{fig:share_district}. The 2023 correlation is 0.011, and classification shows all four poverty by trend quadrants populated, which argues for models that control for unit and year effects rather than simple bivariate diagnostics \cite{seabold2010}. Fourth, the at-risk screen identifies 18 districts with combined high poverty and recent declines, including Clovis, Hobbs, and Zuni. These findings justify targeted stabilization and monitoring while pursuing structural investments in high-need districts \cite{edlawcenter2024,nmped2024}.
\subsection{Extremes: Biggest Movers}
$\,$

$\,$
The largest declines in the latest year include Ruidoso Municipal Schools (\$-9.35\%\$, poverty \textbf{9.4\%}), Dulce Independent Schools (\$-9.13\%\$, \textbf{18.9\%}), Zuni Public Schools (\$-8.57\%\$, \textbf{56.1\%}), Fort Sumner Municipal Schools (\$-8.11\%\$, \textbf{17.0\%}), and Clovis Municipal Schools (\$-7.87\%\$, \textbf{36.8\%}). The largest gains include Belen Consolidated Schools (\$+10.99\%\$, \textbf{18.0\%}), Santa Fe Public Schools (\$+10.36\%\$, \textbf{54.5\%}), Raton Public Schools (\$+8.91\%\$, \textbf{36.4\%}), Moriarty Municipal Schools (\$+7.74\%\$, \textbf{19.1\%}), and Cuba Independent Schools (\$+5.87\%\$, \textbf{18.7\%}). Santa Fe and Zuni both have very high poverty yet move in opposite directions, which shows local dynamics matter.

\begin{figure}[htbp]
\centering
\includegraphics[width=0.8\linewidth]{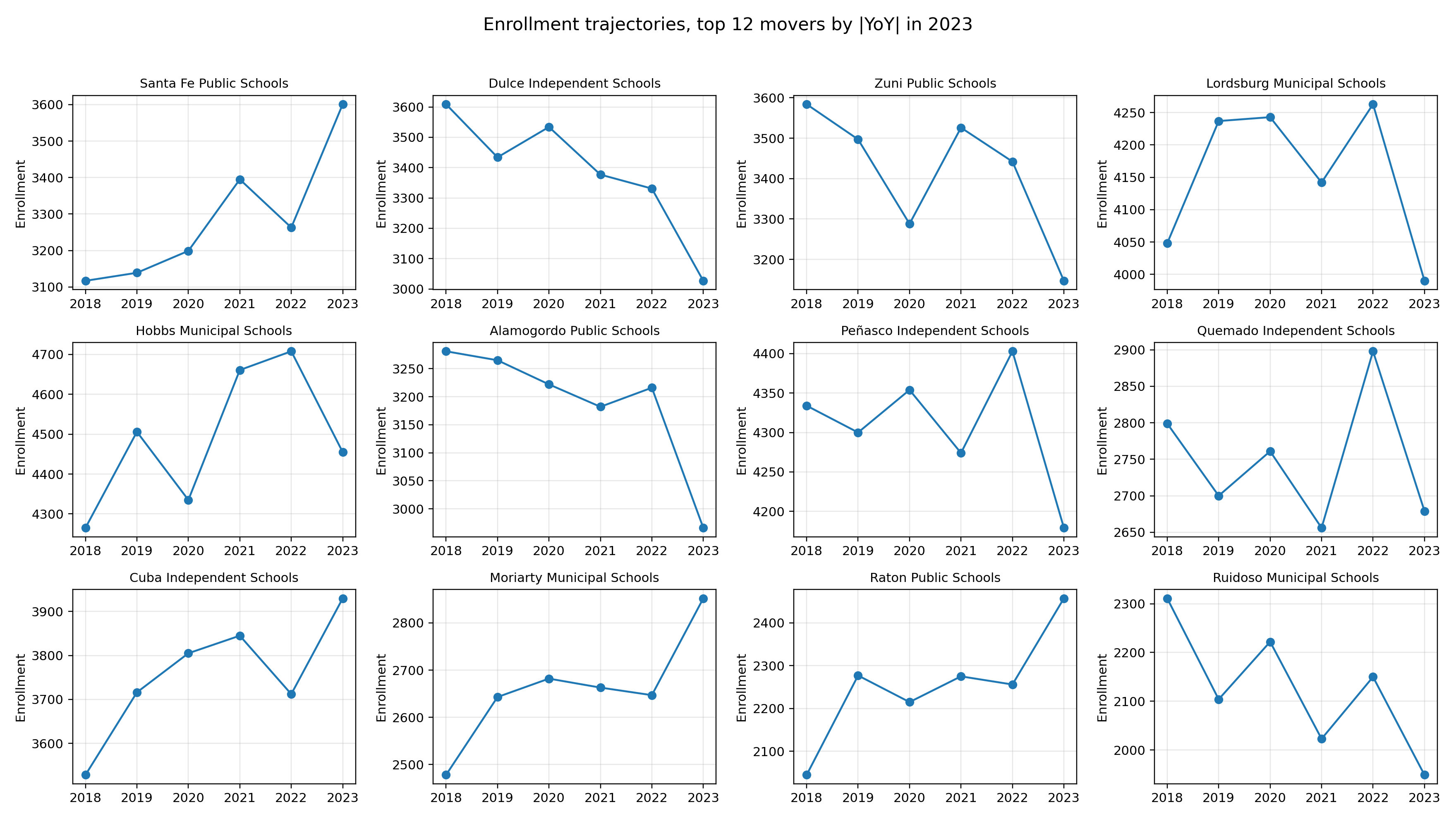}
\caption{New Mexico District Enrollment Trends: Twelve Biggest 2023 Movers}
\label{fig:small_multiples}
\end{figure}

From Figure~\ref{fig:small_multiples} The small-multiple panels highlight sharp 2023 swings among New Mexico’s largest movers. Volatility concentrates in smaller, rural districts serving low-income communities: Zuni declines while Santa Fe rises, illustrating that poverty alone does not determine direction. Multi-year trajectories show jolts rather than smooth trends, consistent with housing shifts, industry cycles, and program changes.
\subsection{Spatial Patterns}
$\,$

$\,$
Maps for 2023 show clusters of high-poverty districts in rural and border-adjacent regions and a checkerboard of YoY changes rather than a single gradient. Neighboring districts with similar poverty can diverge, consistent with localized shocks such as industry cycles, housing availability, and program changes. Population-weighted tract indicators integrated into a single geometry vintage help ensure these contrasts reflect lived context rather than boundary artifacts \cite{census_tiger_2023,census_acs_2023}.
\begin{figure}[htbp]
\centering
\includegraphics[width=0.8\linewidth]{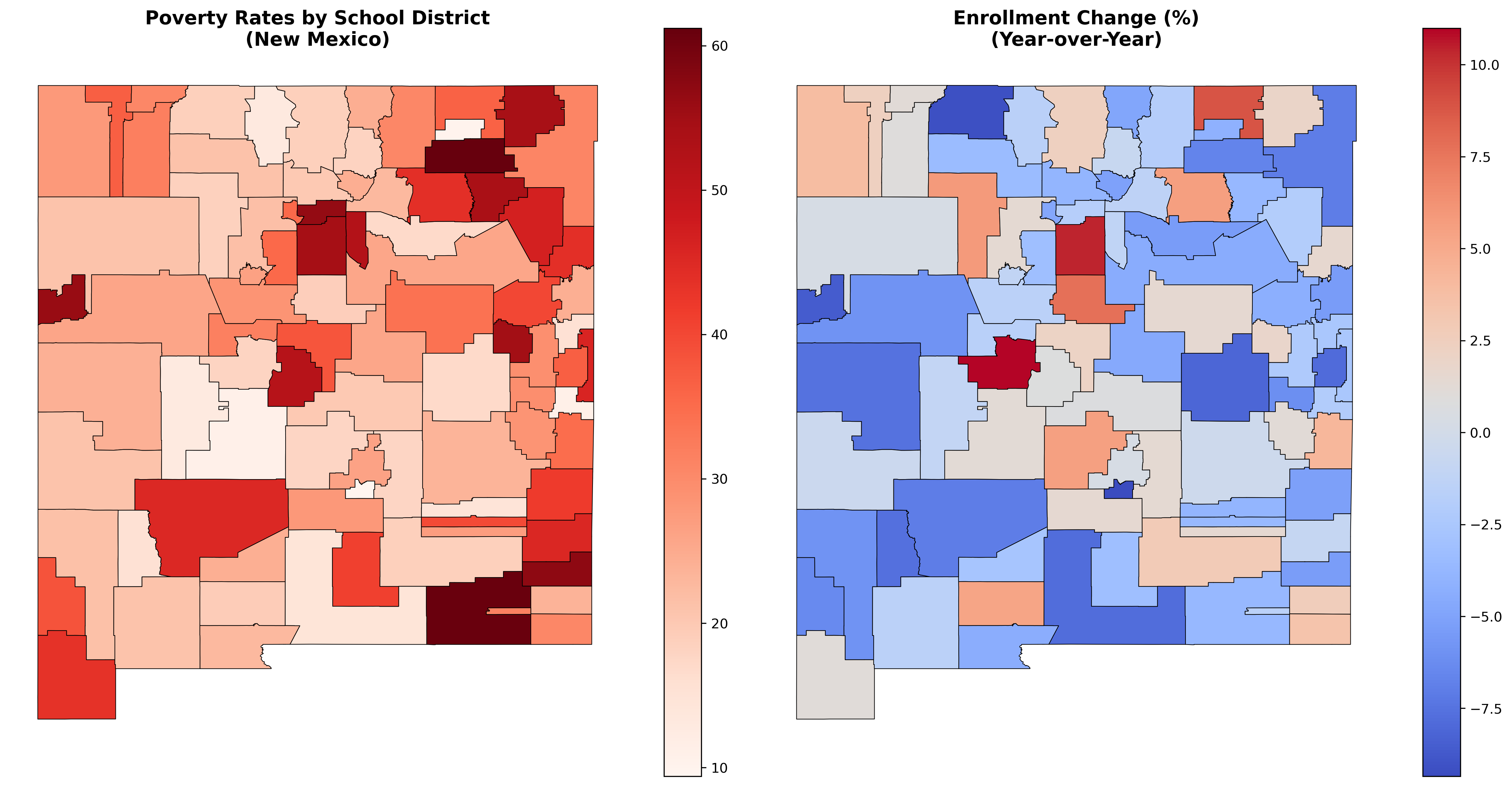}
\caption{New Mexico School Districts (2023): Poverty Intensity vs. Year-over-Year Enrollment Change}
\label{fig:spatial_analysis}
\end{figure}
The Figure~\ref{fig:spatial_analysis} side-by-side maps underscore a weak spatial coupling between poverty intensity and short-run enrollment change. High-poverty clusters appear across rural and border regions, yet adjacent districts show opposite year-over-year shifts—some modest gains amid deep need, others declines in lower-poverty areas. These checkerboard outcomes point to local drivers: housing churn, labor cycles, program mix, and access.
\subsection{Risk Stratification and Early-Warning Signals}
$\,$

$\,$
A composite of poverty intensity and enrollment dynamics flags \textbf{18} \textit{High-risk} districts for immediate attention. Operational screening identifies an early-warning subset of \textbf{53 high-} and \textbf{7 critical-priority} priority labels and a resource-allocation screen that recommends interventions for \textbf{43} districts \textbf{(27 High, 16 Critical)}. Illustrative Critical cases include \textit{Bernalillo}, \textit{Carlsbad}, \textit{Clayton}, \textit{Cloudcroft}, and \textit{Clovis}. Stabilization resources can be sequenced by this watch list while maintaining support for high-poverty districts that are stable or growing \cite{edlawcenter2024,nmped2024}.
\subsection{Policy Implications}
$\,$

$\,$
Two tracks follow. Stabilization should direct flexible funds and student supports to the \textbf{18} high-risk districts and monitor with fast indicators such as attendance and mid-year counts \cite{nmped2024}. Structural investment should recognize concentrated need and enable program breadth, transportation, and staffing pipelines so high-need districts can grow or hold steady \cite{edlawcenter2024,census_acs_2023}. The weak statewide poverty change link underscores the role of local levers rather than baseline need alone.
\section{Conclusion }
$\,$

$\,$
New Mexico's enrollment landscape presents a complex mosaic rather than a uniform trend. Following a period of relative stability from 2018 to 2022, total enrollment declined to 240{,}510 students in 2023, while the average school-age poverty rate persisted near 30 percent \cite{nmped2024,saipe2024}. The near-zero correlation ($r = 0.011$) between poverty levels and year-over-year enrollment changes demonstrates that poverty intensity alone does not determine enrollment trajectories \cite{Smith2022}. Districts are distributed across all four poverty-by-trend categories, with eighteen school systems identified as high-risk due to the convergence of recent enrollment declines and elevated poverty levels \cite{edlawcenter2024}.
These findings support a dual-track policy response. First, immediate stabilization efforts should prioritize high-risk districts through flexible funding mechanisms, comprehensive attendance supports, improved transportation access, and strengthened teacher recruitment pipelines \cite{darling2021}. Second, long-term structural investments must address concentrated disadvantage and enhance educational program accessibility in rural and border regions where geographic isolation compounds socioeconomic challenges \cite{showalter2019}.
Methodologically, the fixed-effects framework with district and year effects and clustered standard errors remains appropriate for ongoing monitoring of enrollment dynamics \cite{seabold2010}. The continued integration of data from the Common Core of Data (CCD), Small Area Income and Poverty Estimates (SAIPE), American Community Survey (ACS), and New Mexico Public Education Department (PED) will maintain a current and actionable panel dataset \cite{nces2024,census_acs_2023}. This integrated approach enables targeted, place-specific interventions that recognize the diverse educational landscapes across New Mexico's school districts.

\bigskip

$\,$

$\,$

\end{document}